\documentclass[aps,prl,twocolumn,groupedaddress,showpacs]{revtex4}

\usepackage{graphicx}
\begin{document}

\title{Spin-spin cross-relaxation in single-molecule magnets}

\author{W. Wernsdorfer$^1$, S. Bhaduri$^2$, R. Tiron$^1$, D.N. 
Hendrickson$^3$, G. Christou$^2$}

\affiliation{
$^1$Lab. L. N\'eel, associ\'e \`a l'UJF, CNRS, BP 166,
38042 Grenoble Cedex 9, France\\
$^2$Dept. of Chemistry, Uni. of Florida, 
Gainesville, Florida 32611-7200, USA\\
$^3$Dept. of Chemistry and Biochemistry, Uni. of California at San 
Diego, La Jolla, California 92093-0358, USA
}

\date{\today}

\begin{abstract}
It is shown that dipolar and weak superexchange interactions 
between the spin systems of single-molecule magnets (SMM) 
play an important role in the relaxation of magnetization. 
These interactions can reduce or increase resonant tunneling. 
At certain external fields, the mechanism of spin-spin 
cross-relaxation (SSCR) can lead to quantum resonances which 
can show up in hysteresis loop measurements as well defined steps. 
A Mn$_4$ SMM is used as a model system to study the SSCR which 
plays also an important role for other SMMs like Mn$_{12}$ or Fe$_8$.
\end{abstract}


\maketitle

Single-molecule magnets (SMMs)~\cite{Sessoli93b,Sessoli93} are one 
of the best systems for studying quantum tunneling 
of large moments~\cite{QTM94}. Each molecule functions as a nanoscale, 
single-domain magnetic particle that, below its blocking 
temperature, exhibits the classical macroscale property 
of a magnet, namely magnetization hysteresis. 
Such a molecule straddles the classical/quantum 
interface in also displaying quantum tunneling of 
magnetization~\cite{Friedman96,Thomas96,Sangregorio97,Aubin98,
Caneschi99,Price99,Yoo_Jae00} and quantum phase interference
~\cite{WW_Science99,WW_JAP02,Garg93}. 
A quantitative understanding 
of the mechanism of magnetization tunneling 
is being developed.
For example, the width of tunnel transitions are 
in general larger than expected from dipolar 
interactions. Crystal defects may play an 
important role: loss or disorder of solvent 
molecules, and even dislocations have been 
proposed~\cite{Chudnovsky01}. 

Since SMMs occur as assemblies in crystals, there is the 
possibility of a small electronic interaction
of adjacent molecules. This leads to very 
small superexchange interactions (or 
exchange interactions, for short) that depend strongly on the 
distance and the non-magnetic atoms 
in the exchange pathway. 
Up to now, such an intermolecular exchange interaction 
has been assumed to be negligibly small. However, our 
recent studies on several SMMs suggest 
that in most SMMs exchange interactions 
lead to a significant influence on the tunnel process. 
Recently, this intermolecular exchange 
interaction was used to couple antiferromagnetically 
two SMMs, each acting as a bias on its neighbor, 
resulting in quantum behavior different from 
that of individual SMMs~\cite{WW_Nature02}. 

The main difference between dipole and exchange 
interactions are: (i) dipole interactions are long 
range whereas exchange interactions are usually short range; 
(ii) exchange interactions can be much stronger 
than dipolar interactions; (iii) whereas the sign 
of a dipolar interaction can be determined easily, 
that of exchange depends strongly on electronic 
details and is very difficult to predict; and
(iv) dipolar interactions depend strongly on the 
spin ground state $S$, whereas exchange interactions 
depend strongly on the single-ion spin states. 
For example, intermolecular dipolar interactions 
can be neglected for antiferromagnetic SMMs 
with $S = 0$, whereas intermolecular exchange 
interactions can still be important and 
act as a source of decoherence. 

In this letter we show that dipolar and/or 
exchange interactions can lead to collective 
quantum processes. The one-body tunnel picture
of SMMs is not always sufficient to explain the
measured tunnel transitions. We propose to improve
the picture by including also two-body tunnel 
transitions such as
spin-spin cross-relaxation (SSCR)~\cite{Bloembergen59,Lukac89}. 
A simple model allows us to explain 
quantitatively all observed transitions.
Including three-body transitions or dealing with
the many-body problem is beyond the slope of
this paper.

The SMM has the formula 
[Mn$_4$O$_3$(OSiMe$_3$)(OAc)$_3$(dbm)$_3$], 
called briefly Mn$_4$. The preparation, 
X-ray structure, and detailed physical characterization 
have been reported~\cite{Bhaduri01}. 
Mn$_4$ crystallizes in a hexagonal space group 
with crystallographic C$_{\rm 3}$ symmetry. 
The complex has a trigonal-pyramidal 
(highly distorted cubane-like) core geometry and 
is mixed-valent: Mn$_3^{\rm III}$Mn$^{\rm IV}$. 
The C$_{\rm 3}$ axis passes through the Mn$^{\rm IV}$ ion 
and the triply bridging siloxide group. 
DC and AC magnetic susceptibility measurements 
indicate a well isolated $S = 9/2$ ground state~\cite{Bhaduri01}. 

All measurements were performed using an 
array of micro-SQUIDs~\cite{WW_PRL99}. 
The high sensitivity  allows us to study single 
crystals of SMMs of the order of 10 to 500 $\mu$m. 
The field can be applied in any direction by separately 
driving three orthogonal coils.

We first review briefly the single spin model 
which is the simplest model describing the spin 
system of an isolated SMM. The spin Hamiltonian is
\begin{equation}
	\mathcal{H}_i = -D S_{z,i}^2 + \mathcal{H}_{{\rm trans}, i} 
	+ g \mu_{\rm B} \mu_0 \vec{S_i}\cdot\vec{H} 
\label{eq_H}
\end{equation}
$S_{x,i}$, $S_{y,i}$, and $S_{z,i}$ are the
components of the spin operator; 
$D$ is the anisotropy constant defining an 
Ising type of anisotropy; $\mathcal{H}_{{\rm trans}, i}$, 
containing $S_{x,i}$ or $S_{y,i}$ spin operators, 
gives the transverse anisotropy which is small 
compared to $D S_{z,i}^2$ in SMMs; and the last term 
describes the Zeeman energy associated 
with an applied field $\vec{H}$. 
The index $i$ labels different SMMs (see below). 
This Hamiltonian has an energy level spectrum 
with $(2S+1)$ values which, to a first approximation, 
can be labeled by the quantum numbers $m = -S, -(S-1), ...,  S$ 
taking the $z$-axis as the quantization axis. 
The energy spectrum can be obtained by using standard 
diagonalization techniques (see Fig. 1). 
At $\vec{H} = 0$, the levels $m = \pm S$ have the 
lowest energy. When a field $H_z$ is applied, 
the levels with $m > 0$ decrease in energy, 
while those with $m < 0$ increase. Therefore, 
energy levels of positive and negative quantum numbers 
cross at certain values of $H_z$ given 
by $\mu_0 H_z \approx n D/g \mu_{\rm B}$, where $n = 0, 1, 2, 3, ...$. 

When the spin Hamiltonian contains transverse terms 
($\mathcal{H}_{\rm trans}$), the level crossings  can 
be {\it avoided level crossings}. The spin $S$ is {\it in resonance} 
between two states when the local longitudinal 
field is close to an avoided level crossing. 
The energy gap, the so-called {\it tunnel splitting}
$\Delta$, can be tuned by a transverse field 
(a field applied perpendicular to the $S_z$ direction) 
via the $S_xH_x$ and $S_yH_y$ Zeeman terms.

The effect of these avoided level crossings 
can be seen in hysteresis loop measurements. 
Figs. 2 and 3 show typical hysteresis loops 
for a single crystal of Mn$_4$. 
When the applied field is near an avoided level 
crossing, the magnetization relaxes faster, 
yielding steps separated by plateaus. 
A closer examination of the tunnel transitions 
however shows fine structures which cannot 
be explained by the above model. We suggest in 
the following that these additional steps are due 
to a collective quantum process, called spin-spin cross-relaxation (SSCR), 
involving pairs of SMMs which are coupled by dipolar 
and/or exchange interactions. Such SSCR processes were recently 
observed in the thermally activated regime of a LiYF$_4$ 
single crystal doped with Ho ions~\cite{Giraud01}.

\begin{figure}
\begin{center}
\includegraphics[width=.46\textwidth]{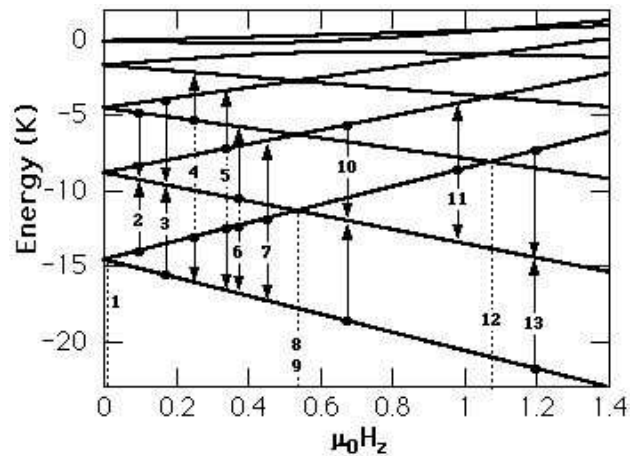}
\caption{Zeeman diagram of the 10 levels of the $S$ = 9/2 
manifold of Mn$_4$ as a function of the field 
applied along the easy axis (Eq. 1 with D = 0.72 K, and a transverse 
anisotropy term  $E (S_x^2 - S_y^2)$ 
with $E$ = 0.033 K~\cite{WW_PRB02}). From bottom to top, 
the levels are labeled with quantum numbers 
m = $\pm9/2, \pm7/2, ..., \pm1/2$. 
The levels cross at fields given by 
$\Delta$$H_z  \approx n \times 0.53$ T, with $n$ = 1, 2, ... 
The arrows, labeled from {\bf 1} to {\bf 13}, indicate 
tunnel transition that are given in Table 1.
}
\label{Zeeman_1mol}
\end{center}
\end{figure}

In order to obtain an approximate understanding 
of the spin-spin cross-relaxation, we consider 
the following Hamiltonian describing two coupled SMMs:
\begin{equation}
	\mathcal{H} = \mathcal{H}_1 + \mathcal{H}_2 +
	J\vec{S_1}\cdot\vec{S_2} 
\label{eq_dimer}
\end{equation}
where each SMM is modeled by a giant spin 
with a spin ground state $S$ and an Ising-like 
anisotropy; the corresponding Hamiltonian 
is given by Eq. 1 where $i$ = 1 or 2 labels the two SMMs. 
The two SMMs are coupled by a small exchange 
interaction $J$ which takes into account 
the contributions of dipole-dipole and/or 
superexchange interactions (Eq. 2). 
The $(2S+1)(2S+1)$ energy states 
of the dimer can be calculated by exact 
diagonalization and are plotted in Fig. 4 
as a function of a magnetic field applied 
along the easy axes of magnetization. 
Any energy level crossing of such a diagram can 
be a possible quantum transition depending 
on the magnitude of transverse terms and
the type of the transition. We will see that 
only few of them are relevant at very low temperatures.

Before proceeding to the detailed discussion 
of this diagram, it is important to note that 
in reality a spin of a SMM is coupled to many 
other SMMs which in turn are coupled to many other 
SMMs. This represents a complicated many-body problem
leading to quantum processes of more than two SMMs. 
However, the more SMMs that are involved, the lower 
is the probability for its occurrence. 
In the limit of small exchange couplings and transverse
terms we propose therefore to consider only processes 
involving one or two SMMs. 
The mutual couplings between all SMMs 
should lead mainly to broadenings and small 
shifts of the observed quantum steps.

\begin{figure}
\begin{center}
\includegraphics[width=.46\textwidth]{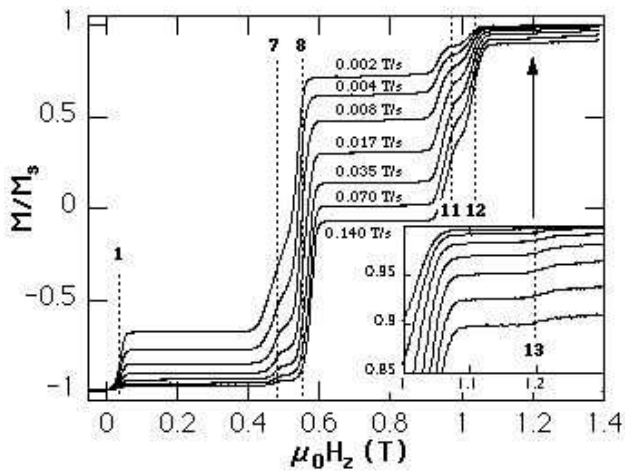}
\caption{Hysteresis loop measurements of a single 
crystal of Mn$_4$ at low temperatures 
(40 mK) where thermal activation to excited spin 
states can be neglected. The field is applied in 
direction of the easy axis of magnetization and 
swept at a constant rate between 0.002 and 0.14 T/s. 
The tunnel transitions are labeled by numbers, see Table 1. 
Inset: Enlargement for the higher field region.
}
\label{hyst_v}
\end{center}
\end{figure}

\begin{figure}
\begin{center}
\includegraphics[width=.46\textwidth]{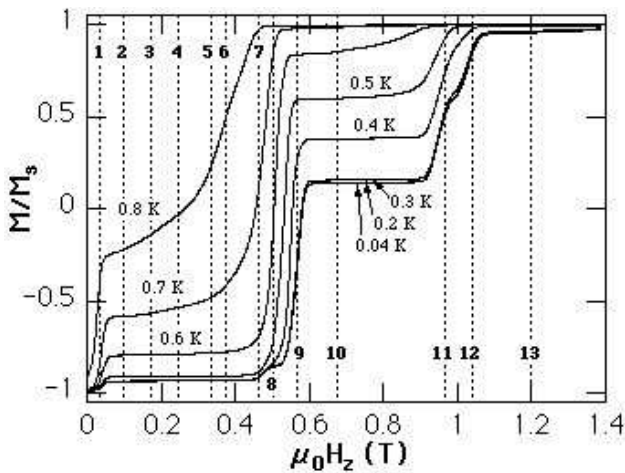}
\caption{Hysteresis loop measurements similar to Fig. 2 
but at different temperatures and for a field sweep 
rate of 0.035 T/s. The tunnel transitions are labeled 
by numbers given in Table 1.
}
\label{hyst_T}
\end{center}
\end{figure}
 
We measured the interactions between 
molecules by using relation measurements as a function
of intitial magnetization and
the {\it hole digging} method~\cite{WW_PRL99,WW_PRL00}. 
We found a fine structure of three in
the zero field resonance that is due to 
the strongest nearest neighbor 
interactions of about 0.036 T along the c-axis 
of the crystals. This coincides with the shortest Mn--Mn
separations of 8.032 A between two molecules along the c-axis, 
while the shortest Mn--Mn separations 
perpendicular to the c-axis are 16.925 A.
We cannot explain the value of 
0.036 T by taking into account only dipolar 
interactions, which should not be larger 
than about 0.01 T. We believe therefore that small 
exchange interactions are responsible for the observed value.
Indeed, the SMMs are held together by two H bonds C--H--O
wich are probably responsible for the small exchange interactions.

We selected 13 levels crossings (see Figs. 1 - 4 and Table 1) 
which we divide into different types and into 
two regimes: (i) the very low temperature regime and 
(ii) the regime of small thermal activation to the 
first activated energy levels. 
In the very low temperature regime, we can neglect 
any activation to excited states. Transition {\bf 1} corresponds 
to the ground state (GS) tunneling from (-9/2,-9/2) to (-9/2,9/2), 
i.e. one of the two coupled spins reverses. 
The coupling to its neighbor leads to a field shift 
of about 0.03 T. Transitions {\bf 8}, {\bf 9}, and {\bf 12}, 
correspond to GS to excited state (ES) tunneling. 

Transition {\bf 7} is a SSCR wherein a pair
of SMMs tunnels from the GS (-9/2,-9/2) to 
the ES (-7/2,9/2). That means that this common 
tunnel transition reverses one of the two spins, 
and the other makes a transition to an excited state. 
This excited state is stable only for a short 
time and relaxes to the GS (-9/2,9/2). 
Transition {\bf 11} is analogous but from 
the GS (-9/2,-9/2) to the ES (-7/2,7/2). 
Transition 13 is again a SSCR but from the 
GS (-9/2,9/2), that is where one spin is already reversed, 
to the ES (7/2,7/2).

Transitions {\bf 2} - {\bf 6}, and {\bf 10} are 
excited state spin-spin cross-relaxations (ES-SSCR); 
that means they reverse from one ES to another ES. 
For example, transition {\bf 10} corresponds to a tunneling 
from (-7/2,9/2) to (7/2,7/2).

The SSCR transitions can be seen as virtual 
phonon transitions. Indeed, whenever there is 
a field where the energy difference between a 
lower lying energy state is equal to that of a higher 
lying state (see Fig. 1), a transition involving two 
SMMs can occur provided that both spins are coupled. 
The transverse terms of the coupling interaction 
produce a tunnel splitting between two coherently 
coupled quantum states. When sweeping the field 
through such a tunnel splitting, there is a Landau-Zener 
tunnel probability of mutual spin flips: one  
molecule transfers to a lower energy state, 
the other to a higher one. The virtual phonon 
transition picture allows one to immediately locate 
possible SSCRs in the single-spin Zeeman diagram (see Fig. 1). 
This method is therefore particularly helpful for large spins 
where an exact diagonalization of the Hamiltonian matrix of the 
coupled SMMs is tedious.

\begin{figure}
\begin{center}
\includegraphics[width=.46\textwidth]{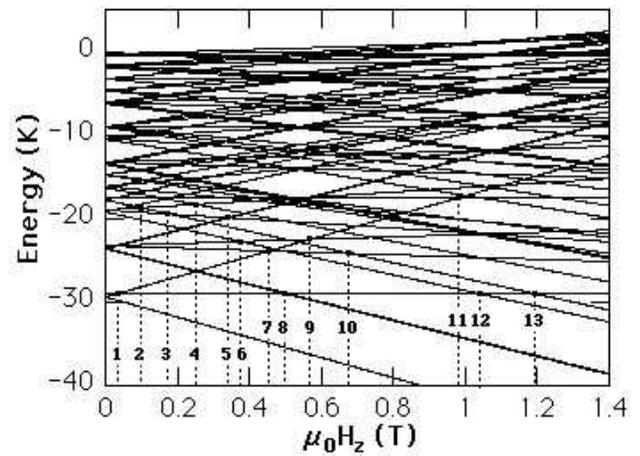}
\caption{The 100 spin state energies of two 
coupled spins $S$ = 9/2, having the same 
anisotropy as in Fig. 1, as a function of 
applied magnetic field (Eq. 2 with $J$ = -0.01 K). 
Dotted lines, labeled {\bf 1} to {\bf 13}, indicate the observed 
tunnel resonances given in Tab. 1.
}
\label{Zeeman_2mol}
\end{center}
\end{figure}

\begin{figure}
\begin{center}
\includegraphics[width=.46\textwidth]{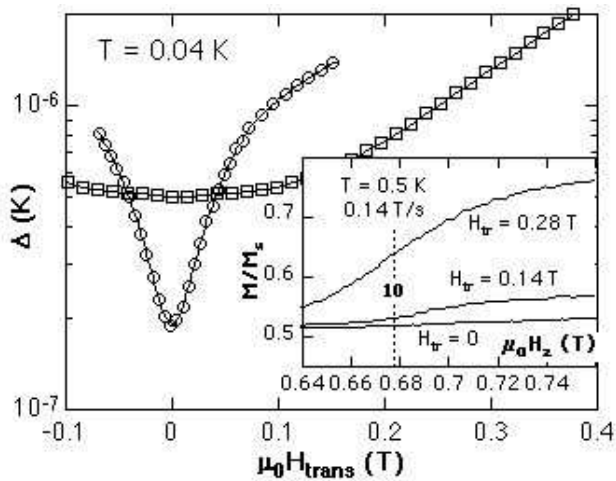}
\caption{Tunnel splitting as a function of transverse 
field for a single molecule transition {\bf 1} (circles) 
and a spin-spin cross relaxation transition {\bf 7} (squares). 
The parity of the involved wave function is established 
by the fact that transition {\bf 1} is very sensitive to a 
transverse field (odd transition) whereas transition {\bf 7} 
depends only smoothly on the transverse field (even transition).
Inset: Enlargement of hysteresis loop measurements at three 
different transverse fields showing the ES-SSCR transition {\bf 10}.
}
\label{delta}
\end{center}
\end{figure}

We checked that all transitions {\bf 1} - {\bf 13}
are sensitive to an applied transverse field, which
always increases the tunnel rate. The inset of Fig. 5 
presents a typical example showing the
transverse field dependence of the ES-SSCR transition {\bf 10}.
Fig. 5 presents a measurement of the
tunnel splitting of transition {\bf 1} and transition {\bf 7}
using the Landau--Zener method~\cite{WW_Science99} 
which establishes also the parity
of the avoided level crossing~\cite{WW_PRB02}.
In between the transitions {\bf 1} - {\bf 13}, other
avoided level crossings can be found (Fig. 4) that requires 
both SMMs to tunnel simultaneously. The corresponding
tunnelling probability is much small and will be discussed 
elsewhere.

In the low-temperature regime, the strongest 
observed SSCR concerns 
the transitions {\bf 7} and {\bf 11}. The question arises 
whether such transitions also play a role in 
other SMMs like Fe$_8$ and Mn$_{12}$. 
A diagonalization of the spin-Hamiltonian 
of such molecules shows clearly that 
spin-spin cross-relaxation should occur. 
However, it turns out that these transitions 
are very close to the single spin tunnel 
transitions and only broaden them. 
Thermally excited spin-spin cross-relaxation 
should however be observable and might be 
responsible for the fine structures seen in experiments 
of Bokacheva {\it at al.}~\cite{Kent00b} on Mn$_{12}$ 
and of Gaudin~\cite{Gaudin01} and Wernsdorfer~\cite{WW} on Fe$_8$.

R. Giraud is acknowledged 
for fruitful discussions, and 
C. Thirion, D. Mailly and A. Benoit for 
support to the micro-SQUID technique.
This work was supported by the 
European Union TMR network MOLNANOMAG, 
HPRN-CT-1999-0012 and the USA National 
Science Foundation, DMR-0103290.

\begin{table}
\caption{The 13 tunnel transitions, which are 
labeled from 1 to 13 in Figs. 1 - 4, for two coupled SMMs
with S = 9/2. 
Their states are labeled by two quantum numbers 
($m_1$,$m_2$) where $m_i$ = -9/2, -7/2, ..., 9/2. 
For clarity, degenerate states such as ($m$,$m'$) 
and ($m'$,$m$) are not both listed. 
IS: Initial state; FS: Final state; 
GS: ground state; ES: excited state; 
SSCR: spin-spin cross-relaxation. 
\label{tab1}}
\begin{tabular}{|c|c|c|c|}
\hline
n$^{\circ}$	& IS &	FS	& Type\\
\hline
1 &	(-9/2,-9/2) &	(-9/2,9/2) &	GS-GS \\
2 &	(-9/2,5/2) &	(7/2,7/2) &	ES-SSCR  \\	
3 &	(-5/2,9/2) &	(7/2,7/2) &	ES-SSCR  \\	
4 &	(-9/2,5/2) &	(9/2,3/2) &	ES-SSCR  \\	
5 &	(-9/2,-7/2) &	(9/2,-5/2) &	ES-SSCR  \\	
6 &	(-9/2,7/2) &	(9/2,5/2) &	ES-SSCR  \\	
7 &	(-9/2,-9/2) &	(-7/2,9/2) &	SSCR \\
8 &	(-9/2,9/2) &	(9/2,7/2) &	GS-ES  \\
9 &	(-9/2,-9/2) &	(-9/2,7/2) &	GS-ES  \\
10 &	(-7/2,9/2) &	(7/2,7/2) &	ES-SSCR  \\	
11 &	(-9/2,-9/2) &	(-7/2,7/2) & SSCR \\
12 &	(-9/2,9/2) &	(9/2,5/2) &	GS-ES  \\ 	
13 &	(-9/2,9/2) &	(7/2,7/2) &	SSCR  \\	
\hline
\end{tabular} \\
\end{table}


\begin{thebibliography}{10}

\bibitem{Sessoli93b}
R. Sessoli, H.-L. Tsai, A.~R. Schake, S. Wang, J.~B. Vincent, K. Folting, D.
  Gatteschi, G. Christou, and D.~N. Hendrickson, J. Am. Chem. Soc. {\bf 115},
  1804  (1993).

\bibitem{Sessoli93}
R. Sessoli, D. Gatteschi, A. Caneschi, and M.~A. Novak, Nature {\bf 365},  141
  (1993).

\bibitem{QTM94}
{\em Quantum Tunneling of Magnetization-QTM'94}, Vol.~301 of {\em NATO ASI
  Series E: Applied Sciences}, edited by L. Gunther and B. Barbara (Kluwer
  Academic Publishers, London, 1995).

\bibitem{Friedman96}
J.~R. Friedman, M.~P. Sarachik, J. Tejada, and R. Ziolo, Phys. Rev. Lett. {\bf
  76},  3830  (1996).

\bibitem{Thomas96}
L. Thomas, F. Lionti, R. Ballou, D. Gatteschi, R. Sessoli, and B. Barbara,
  Nature (London) {\bf 383},  145  (1996).

\bibitem{Sangregorio97}
C. Sangregorio, T. Ohm, C. Paulsen, R. Sessoli, and D. Gatteschi, Phys. Rev.
  Lett. {\bf 78},  4645  (1997).

\bibitem{Aubin98}
S.~M.~J. Aubin, N.~R. Dilley, M.~B. Wemple, G. Christou, and D.~N. Hendrickson,
  J. Am. Chem. Soc. {\bf 120},  839  (1998).

\bibitem{Caneschi99}
A. Caneschi, D. Gatteschi, C. Sangregorio, R. Sessoli, L. Sorace, A. Cornia,
  M.~A. Novak, C. Paulsen, and W. Wernsdorfer, J. Magn. Magn. Mat. {\bf 200},
  182  (1999).

\bibitem{Price99}
D.~J. Price, F. Lionti, R. Ballou, P.T. Wood, and A.~K. Powell, Phil. Trans. R.
  Soc. Lond. A {\bf 357},  3099  (1999).

\bibitem{Yoo_Jae00}
J. Yoo, E.~K. Brechin, A. Yamaguchi, M. Nakano, J.~C. Huffman, A.L. Maniero,
  L.-C. Brunel, K. Awaga, H. Ishimoto, G. Christou, and D.~N. Hendrickson,
  Inorg. Chem. {\bf 39},  3615  (2000).

\bibitem{WW_Science99}
W.Wernsdorfer and R. Sessoli, Science {\bf 284},  133  (1999).

\bibitem{WW_JAP02}
W. Wernsdorfer, M. Soler, G. Christou, and D.N. Hendrickson, J. Appl. Phys.
  {\bf 1},  1  (2002).

\bibitem{Garg93}
A. Garg, EuroPhys. Lett. {\bf 22},  205  (1993).

\bibitem{Chudnovsky01}
E.~M. Chudnovsky and D.~A. Garanin, cond-mat/0105195 {\bf 87},  187203  (2001).

\bibitem{WW_Nature02}
W. Wernsdorfer, N. Aliaga-Alcalde, D.N. Hendrickson, and G. Christou, Nature
  {\bf 416},  406  (2002).

\bibitem{Bloembergen59}
N.~Bloembergen et~al., Phys. Rev. {\bf 114},  445  (1959).

\bibitem{Lukac89}
M. Lukac, F.~W. Otto, and E.~L. Hahn, Phys. Rev. A {\bf 39},  1123  (1989).

\bibitem{Bhaduri01}
S. Bhaduri, M. Pink, K. Folting, W. Wernsdorfer, D.N. Hendrickson, 
and G. Christou, manuscript in preparation.

\bibitem{WW_PRL99}
W. Wernsdorfer, T. Ohm, C. Sangregorio, R. Sessoli, D. Mailly, and C. Paulsen,
  Phys. Rev. Lett. {\bf 82},  3903  (1999).

\bibitem{Giraud01}
R. Giraud, W. Wernsdorfer, A.~M. Tkachuk, D. Mailly, and B. Barbara, Phys. Rev.
  Lett. {\bf 87},  057203  (2001).

\bibitem{WW_PRL00}
W. Wernsdorfer, A. Caneschi, R. Sessoli, D. Gatteschi, A. Cornia, V. Villar,
  and C. Paulsen, Phys. Rev. Lett. {\bf 84},  2965  (2000).

\bibitem{WW_PRB02}
W. Wernsdorfer, S. Bhaduri, C. Boskovic, G. Christou, and D.N. Hendrickson,
  Phys. Rev. B {\bf 1},  1  (2002).

\bibitem{Kent00b}
L. Bokacheva, A.D. Kent, and M.A. Walters, Phys. Rev. Lett. {\bf 85},  4803
  (2000).

\bibitem{Gaudin01}
G. Gaudin, Ph.D. thesis, Joseph Fourier University, Grenoble, 2001.

\bibitem{WW}
W. Wernsdorfer et al., unpublished.

\end{thebibliography}
\end{document}